\documentstyle[twocolumn]{article}
\begin{document}
\pagestyle{empty}

~
\vspace*{1.5cm}

\twocolumn[
\vspace*{1.5cm}

\noindent {\Large Determination of the strong g-coupling and of
the $\mu_{B^*B\gamma}-$ strength from two-photon decays of heavy
vector mesons}\\
\
\\ D. Guetta$^a$, and P. Singer$^a$\\ \ \\
$^a$
Department of Physics, Technion--Israel Institute of Technology,
Haifa 32000, Israel \\
\\ \ \\

We describe a calculation of the two-photon decays of heavy
vector-mesons, $B^*\rightarrow B\gamma\gamma$ and $D^*\rightarrow
D\gamma\gamma$, by using the ``heavy meson chiral Lagrangian''.
The decay amplitudes are expressed in terms of the strong coupling
$g$ of the Lagrangian at various powers and the strength of the
anomalous magnetic dipole $\mu_{B^*B\gamma}$ (respectively
$\mu_{D^*D}\gamma$). In the charm case we are able to express the
branching ratio $\Gamma(D^{*0}\rightarrow
D^0\gamma\gamma)/\Gamma(D^{*0})$ as a function of $g$ only and we
expect it to be in the $10^{-6}-10^{-5}$ range, depending on the
value of $g$.  The determination of
$\mu_{B^{*0}B^{0}\gamma},\mu_{B^{*+}B^+}\gamma$ requires a more
involved analysis, including the consideration of bremsstrahlung
radiation for the $B^{*+}\rightarrow B^+\gamma\gamma$ case.\\ \ \\
\ \\]

\noindent{\bf 1. INTRODUCTION }\\

The heavy vector mesons $D^*$ and $B^*$ decay by strong and
electromagnetic interactions. Since the mass difference
$M_{B^*}-M_B$ is only $45.78\pm 0.35$ MeV [1], the decay
$B^*\rightarrow B\pi$ cannot occur and the main decay of $B^*$ is
$B^*\rightarrow B\gamma$. On the other hand, the $D^*-D$ mass
difference barely allows the decay to a pion and accordingly the
main decays are $D^*\rightarrow D\pi, D\gamma$. The strength of
the couplings involved in these processes, $g_{D^*D\pi}$,
$g_{D^*D\gamma}$, $g_{B^*B\gamma}$ for the different charge states
is not known although relative branching ratios for the various
$D^*\rightarrow D\pi$, $D^*\rightarrow D\gamma$ isotopic spin
states are well measured [1].

In recent years, a theoretical framework has been developed for
the description of the low energy strong interactions [2,3] and of
the electromagnetic interactions [3,4] between mesons containing
a heavy quark $Q$ and the pseudoscalar Goldstone bosons. The Heavy
Quark Chiral Lagrangian (HQ$\chi$L) developed for this framework
[2,3] combines the flavour and spin symmetry of the heavy quark
effective theory
 with the $SU(3)_L \otimes SU(3)_R$ chiral
symmetry of the light sector, allowing the use of chiral
perturbation theory for treating the interaction of heavy mesons
with low-momentum pions. The symmetry of this Lagrangian leads to
defining a strong coupling $g$ which represents the four couplings
$g_{B^*B\pi}$, $g_{B^*B^*\pi}$, $g_{D^*D^*\pi}$, $g_{D^*D\pi}$ (as
well as couplings to $\eta$, etc) and a coupling $\mu$ which
represents the anomalous magnetic couplings $g_{B^*B^*\gamma}$,
$g_{B^* B\gamma}$, $g_{D^*D^*\gamma}$, $g_{D^*D\gamma}$. This is true in the symmetry
limit and corrections will obviously appear when one considers
deviations from both the heavy quark limit and the zero-mass
chiral limit.

At present, there is no experimental determination of the
strength of any of these couplings. On the other hand a variety of
theoretical models has been advanced for their calculation (see
[5] for a recent review). As a result, there is a rather wide
range of theoretical possibilities for $g$ and $\mu$, since the
predictions of various models differ considerably. In the next
section we shall present a short survey of this situation. One
might expect a direct determination of $g_{D^*D\pi}$ from the
measurement of the decay width of $D^*\rightarrow D\pi$;
presently, there is only an upper
limit of $\Gamma_{\rm
tot}(D^{*+}) < 131$keV [6] and the possibility
of an actual
measurement depends on the magnitude of the physical value.  It is
therefore of obvious interest to devise independent methods for
the determination of $g$ and $\mu$ couplings. We shall describe
here our recent suggestion [7,8] of using the $B^*\rightarrow
B\gamma\gamma$ and $D^*\rightarrow D\gamma\gamma$ decays for the
determination of $g$ and $g_{B^*B\gamma}$.  Using the Heavy Quark
Chiral Lagrangian we shall express the amplitudes for
$B^*\rightarrow B\gamma\gamma$, $D^*\rightarrow D\gamma\gamma$ in
terms of $g$ and $\mu$. A detailed analysis will show that the
$D^*$ decay is particularly useful for the determination of $g$
while the $B^{*+}\rightarrow B^{+}\gamma\gamma$ decay may possibly be used
to determine the strength of $g_{B^{*+}B^+\gamma}$.\\

\noindent{\bf 2. THE THEORETICAL FRAMEWORK}\\

The decays we are considering, which were mentioned firstly in
[7], are $B^{*+}\rightarrow B^+\gamma\gamma$, $B^{*0}\rightarrow
B^0\gamma\gamma$, $B^{*0}_s\rightarrow B^0_s\gamma\gamma$ and
$D^{*+}\rightarrow D^+\gamma\gamma$, $D^{*0}\rightarrow
D^0\gamma\gamma$, $D^{*+}_s\rightarrow D_s^+\gamma\gamma$. We
shall use  for their description the HQ$\chi$L in the leading
order of chiral perturbation theory. Corrections to it requires
terms proportional to the mass of the light quarks
$m_q$, to ($1/M_Q$) and derivatives in the HQ$\chi$ Lagrangian
[9-11]; this question will be commented on in the  last section.

The interaction term of HM$\chi$L to lowest order is [2,3]
\begin{eqnarray}
{\cal L}^{\rm int}=g Tr\left\{ \bar{H}_a\gamma_\mu\gamma_5
A^\mu_{ab}H_b\right\} 
\end{eqnarray}
which defines the basic strong-interaction coupling  $g$. $H_a$ is
a $4\times 4$ Dirac matrix, with one spinor index for  the heavy
quark $Q$ and the other for the light quark $q$,
\begin{eqnarray}
H=\frac{1+ v\hspace{-0.18cm}/}{2}
\left[B^*_\mu\gamma^\mu-B\gamma_5\right], \ \ \bar{H}= \gamma_0
H^\dagger\gamma_0 
\end{eqnarray}
and $B^*_{a\mu}(v)$, $B_a(v)$ are the velocity-dependent
annihilation operators of the meson fields, with
$v^\mu B^*_{a\mu}=0$. $a,b$ denote light quark flavours
$(a,b=1,2,3)$ and the $B^*_\mu$, $B$ fields represent, unless
otherwise specified, both the beauty and charm sectors.
$A^\mu_{ab}$ is the axial current containing an odd number of pion
fields, given by
$A_\mu=\frac{1}{2}(\zeta^\dagger\partial_\mu\zeta-\zeta\partial_\mu\zeta^\dagger)$,
$\zeta = \exp(i{\cal M}/f)$ with ${\cal M}$ being the $3\times 3$
matrix of the octet of pseudoscalar Nambu-Goldstone bosons. $f$ is
the pion decay constant, $f=132$ MeV. Expanding $A_\mu$ and keeping
the first term $A^\mu = (i/f)\partial_\mu{\cal M}$ we obtain the
explicit interaction terms from (1)
\begin{eqnarray}
{\cal L}^{\rm int}_{(1)} =
\left[-\frac{2g}{f} B^*_\mu\partial^\mu{\cal M} B^\dagger+H.c.\right]
\nonumber \\ + \frac{2gi}{f} \epsilon_{\mu\nu\sigma\tau} B^{*\mu}
\partial^\sigma{\cal M} B^{*\dagger\nu}v^\tau \ . 
\end{eqnarray}
With the usual definition for the strong $B^*B\pi$, $B^*B^*\pi$
[5] vertices
\begin{eqnarray}
\langle \pi (q)\bar{B}(v_1)|B^*(v_{2,} \epsilon_2)\rangle =
g_{B^*B\pi}(q^2) q_\mu \epsilon^\mu_2 
\end{eqnarray}
\begin{eqnarray}
\hspace*{-2cm}&& \langle\pi(q)\bar{B}^*(v_1,\epsilon_1)|B^*(v_2,\epsilon_{2})\rangle\nonumber
\\
&&\hspace{1cm} = g_{B^*B^*\pi}(q^2)\epsilon_{\alpha\beta\mu\nu}\epsilon_1^\alpha\epsilon_2^\beta
q^\mu v_1^\nu 
\end{eqnarray}
the physical couplings are given by the limit $q^2\rightarrow
m_\pi^2$. Isospin symmetry is expressed by the relations
\begin{eqnarray}
g_{B^*B\pi} &\equiv& g_{B^{*+}B^0\pi^+} = -
\sqrt{2}g_{B^{*+}B^+\pi^0} =\nonumber \\
&& \sqrt{2} g_{B^{*+}B^0\pi^0} = -
g_{B^{*0}B^+\pi^-} 
\end{eqnarray}
and similarly for the $g_{B^*B^*\pi}$ couplings. Comparing (3) to
(4), (5) we obtain these couplings expressed in terms of $g$, and
at this point we distinguish between beauty and charm
$$
g_{B^*B\pi} = g_{B^*B^*\pi} = (2M_B/f)g \ ,\eqno(7.1)
$$
$$
g_{D^*D\pi} = g_{D^*D^*\pi} = (2M_D/f)g \ . \eqno(7.2)
$$
Throughout this work, we follow the normalization conventions of
[5]. In (7.1), (7.2) mass degeneracy for $(B^*,B)$ and $(D^*,D)$
pairs was assumed and the use of the pseudoscalar mass value in
these equations is conventional.

The incorporation of electromagnetism in HM$\chi$L is performed by
the usual procedure of minimal coupling, which leads to the
replacement of derivatives in the free Lagrangian ${\cal L}_0$ by
covariant derivatives containing the $U(1)$ photon field, when
$$
{\cal L}_0 = - i Tr\left\{ \bar{H}_a v^\mu D_{\mu b a} H_b
\right\} + $$
$$ \frac{f^2}{8} Tr\partial_\mu \Sigma \partial^\mu \Sigma^\dagger \ . \eqno(8)
$$

Here $\Sigma = \zeta^2(X)$ and $D_\mu=\partial_\mu+V_\mu$ with the
vector current, containing an even number of pions, given by
$V_\mu=\frac{1}{2}(\zeta^\dagger\partial_\mu\zeta + \zeta
\partial_\mu\zeta^\dagger)$

The interaction vertices of the heavy mesons with photons are then
[3].
$$ \langle \gamma (k, \epsilon)\bar{B}(v_1)|B(v_2)\rangle = e M_B(v_1+v_2)\cdot \epsilon\eqno(9.1)
$$
$$\langle \gamma (k,\epsilon)\bar{B}^*(v_1,\epsilon_1)|B^*(v_2,\epsilon_2)\rangle
=$$
$$ e M_{B^*}(\epsilon_1\cdot\epsilon_2)(v_1+v_2)\cdot \epsilon \ . \eqno(9.2)
$$
These vertices are not capable to provide for the existing
$B^*B\gamma$ interaction and an additional gauge invariant term
proportional to the electro magnetic field $F_{\mu\nu}$ must be
added [3,4,11] to the Lagrangian ${\cal L}_0 + {\cal L}^{\rm
int}$. It is given as
$$
{\cal L}^{(\mu)} = \frac{e\mu}{4} Tr \{ \bar{H}_a \sigma_{\mu\nu}
F^{\mu\nu} H_b \delta_{ab}) \eqno(10)
$$
which gives when we express $H_a$ from Eq.~(2)
$$
{\cal L}^{(\mu)} = - e\mu F^{\mu\nu} \left[i B^{*+}_\mu B^*_\nu +
\epsilon_{\mu\nu\sigma\tau} v^\sigma(B^+B^{*\tau} + H.c.)\right] \
.
\eqno(11)
$$
This translates into two ``anomalous'' vertices for the
electromagnetic interaction, in addition to (91.), (9.2),
$$
\langle \gamma(k,\epsilon)\bar{B}^*(v_1,
\epsilon_1)|B^*(v_2,\epsilon_2)\rangle =
$$
$$ e\mu
M_{B^*}(\epsilon_1\cdot k \epsilon\cdot \epsilon_2 -
\epsilon_2\cdot k \epsilon \cdot \epsilon_1) \ , \eqno(12.1)
$$
$$
\langle \gamma(k,\epsilon)\bar{B}(v_1)|B^*(v_2,\epsilon_2)\rangle
= $$
$$
- i e \mu M_{B^*}\epsilon_{\mu\nu\sigma\tau} \epsilon^\mu k^\nu
v_2^\sigma \epsilon_2^\tau \ . \eqno(12.2)
$$

The new coupling $\mu$ is related to the customary physical
couplings by
$$M_{B^*}\mu^0 = g_{B^{*0}B^0\gamma}; \ \ M_{D^*}\mu^0 =
g_{D^{*0}D^0\gamma}  \ , \eqno(13.1)
$$
$$M_{B^*}\mu^+ = g_{B^{*+}B^+\gamma}; \ \ M_{D^*}\mu^+ =
g_{D^{*+}D^+\gamma}  \ . \eqno(13.2)
$$
The values of $g_{B^*B\pi}$, $g_{D^*D\pi}$, $g_{B^* B*\pi}$,
$g_{D^*D^*\pi}$, $g_{B^*B\gamma}$, $g_{D^*D\gamma}$,
$g_{B^*B^*\gamma}$, $g_{D^*D^*\gamma}$ are not known from direct
experiments; however, a vast literature of theoretical attempts
has appeared and we expose now succinctly the emerging theoretical
picture concerning $g$ and $\mu$ and the respective physical
couplings expressed in Eqs.~(7.1),(7.2),(13.1),(13.2). It should
be stressed here that the wide interest in $g$ is due also to the
fact that its strength appears in the expressions for many
low-energy electroweak processes, like $B\rightarrow \pi\ell\nu$,
$D_s \rightarrow K\ell\nu$, $B\rightarrow D^*\pi\ell\nu$
 among others (see [5] for
comprehensive review).

One may group roughly the calculations of $g$ into four major
classes, those based on constituent quark models, those using
relativistic quark models, the use of QCD sum rules and the
effective Lagrangian approach. There are obviously different
approaches within the same group as well as calculations which
cannot be classified as mentioned.  Most calculations obtain, in
fact, the values of $g_{B^*B\pi}$ and/or $g_{D^*D\pi}$. The naive
quark model result is $g=1$ [3] and its modification by the
inclusion of chirality [12] brings it slightly down to $g\simeq
0.75-0.8$. The use of relativistic quark approaches [13], of which
recently a new wave of results has appeared gives values of $g$ of
approximately 0.45-0.65. The QCD sum-rules tend to give [14,15]
lower values, $g\simeq 0.2-0.4$. We also mention a lattice QCD
calculation [16] with the result $g=0.42$ (4), the analysis of
Stewart [11] which uses HQ$\chi$L with corrections to order $m_q$ and
$1/M_Q$ and constraints from the $D^*$ decay branching ratio to
obtain $g=0.27^{+0.09}_{-0.04}$, and a chiral bag model
calculation [17] which has predicted accurately the observed
$D^*$ branching ratio and gives $g=0.53$. We should caution the
reader on the use in the literature of different
definitions for the couplings involved. Thus $g_{D^*D\pi}$ defined
in [15] is larger by the factor $(M_{D^*}/M_D)$ than the one used,
e.g. in Refs.~[5,7,12]. Moreover, in certain works one defines a
coupling $\hat{g}$ (see [15]), which is related to $g$ of
Eqs.~(7.1), (7.2) by
$\hat{g}=(M_{D^*}/M_D)\left(1+\frac{\Delta}{M_D}\right)^{-1} g$,
where $\Delta$ is $1/M_Q$ correction with a value of $0.7\pm 0.1$.

Experimentally [6], there is the upper limit of
$\Gamma(D^{*+}\rightarrow D^0\pi^+) < 89$KeV, which translates
into $ g < 0.71$. The prediction of the naive quark model for this
mode is 120-180 KeV, of the relativistic quark models 55-80 KeV,
of the QCD sum rules 10-30 KeV, of Ref.[17] is $53\pm 3$ KeV and
of Ref.~[11] is 18 KeV.

The situation concerning $\mu$ is very similar. Generally, the
same models employed for calculating $g$ were used also for
getting the $D^*\rightarrow D\gamma$, $B^*\rightarrow B\gamma$
modes. Since the relative branching ratios for $D^{*+,0}$ decays
into $D\pi$ vs. $D^{+,0}\gamma$ are known [1], most calculations
insist on reproducing these ratios; then the calculated absolute
partial widths for $D^*\rightarrow D\gamma$ decays follow the same
pattern as the calculations of $g$, i.e. predicting rather small
width from QCD sum rules [14,18] and larger ones from quark models
[13], chiral approaches [4,17] or potential models [19]. The range
of variation for the different predictions is about one order of
magnitude in rate. Parallel predictions are made for the
$B^{*0}\rightarrow B^{0}\gamma$, $B^{*+}\rightarrow B^+\gamma$
decays. To exemplify with typical results, the QCD sum-rule
approach of Dosch and Narrison and of [18] predicts
$\Gamma(B^{*0}\rightarrow B^0\gamma)= 0.04$KeV,
$\Gamma(B^{*+}\rightarrow B^+\gamma)= 0.1$KeV, while a chiral bag
model calculation [20] obtains $\Gamma(B^{*0}\rightarrow B^0\gamma)=
0.28$KeV, $\Gamma(B^{*+}\rightarrow B^{+}\gamma)= 0.62$KeV and
even higher values were obtained in certain models ([4], [9] and
first ref. of [13]).\\

\noindent{\bf 3. TWO PHOTON DECAY AMPLITUDES}\\

We use now HM$\chi$L with electromagnetic interactions, as
detailed in the previous section, to calculate the $B^*\rightarrow
B\gamma\gamma$ process ($B$ stands for both beauty and charm). We
begin by treating the neutral decays only, which are free from
bremsstrahlung. The calculation is performed to leading order in
chiral perturbation theory and to this order there are no unknown
counterterms [11,21]. The decay amplitude is given at this  order
by [7,8]
$$A^0=A^0_{\rm anomaly} + A^0_{\rm tree} + \sum^6_{i=1}
A^{0(i)}_{\rm loop} \ . \eqno(14) $$
We shall describe now these various contributions to the decay
amplitude stressing the couplings, without giving the
involved detailed expressions which can be found in [7].

$A^0_{\rm anomaly}$ represents $B^{*0}\rightarrow
B^{0}$``$\pi$''$\rightarrow B^0\gamma\gamma$ via a virtual pion. In the
charm case, where the decay to a physical pion is allowed, we
limit the $s=(k_1 + k_2)^2$ variable to be up to 20 MeV away from
the pion mass. This contribution is proportional to $\alpha g$,
with $\alpha = e^2/4\pi$. $A^0_{\rm tree}$ is given by the
transition $B^{*0}\rightarrow $``$B^{*0}$''$
\gamma \rightarrow B\gamma\gamma$,
having two insertions of $\mu$. Hence $A^0_{\rm tree}$ is
proportional to $\alpha \mu^2_0$. Then we have six classes of
diagrams containing pions and/or  kaons loop, and the photon
emission may occur from loop or from the external legs via the
anomalous $\mu$ interaction. We shall describe a typical diagram
for each of these six classes. $A^{0(1)}_{\rm loop}$ describes
$B^{*0}\rightarrow(B^{*+}\pi^-)\rightarrow B^0\gamma\gamma$ with
the two photons radiated from the virtual charged pion and is
proportional to $\alpha g^2$. The radiation from the
virtual $B^{*+}$ is negligible. $A^{0(2)}_{\rm loop}$ represents
the transition $B^{*0} \rightarrow $``$B^{*0}$''$\gamma \rightarrow
(B^{*+}\pi^-) \gamma \rightarrow B^0\gamma\gamma$ and is proportional
to $\alpha \mu_0g^2$. $A^{0(3)}_{\rm loop}$ is a similar class,
with the $B^{*0}B^0\gamma$ vertex replacing the
$B^{*0}B^{*0}\gamma$ one in the first step, thus being also
proportional to $\alpha\mu_0g^2$. $A^{0(4)}_{\rm loop}$ describes
the transition $B^{*0}\rightarrow $``$(B^*\pi)$''$\gamma \rightarrow $``$
B^{*0}$ ''$ \gamma \rightarrow B^0\gamma\gamma$ being proportional to
$\alpha g^2\mu_0$. Replacing $B^*$ by $B$ in the loop one gets
$A^{0(5)}_{\rm loop}$ which is also proportional to $\alpha
g^2\mu_0$. Finally, we have $B^{*0}
\rightarrow $``$(B^{*+}\pi^-)$''$\gamma\gamma \rightarrow
B^0\gamma\gamma$; this describes double radiation by the loop,
with one photon radiated by the pion and the other one due to the
anomalous transition $\mu$. This amplitude is proportional to
$\alpha\mu_+g^2$.

Calculating the rate of the decay one obtains an expression
containing 13 terms which depend on the products
$g^\alpha\mu_0^\beta\mu^\gamma_+$ with different powers, $\alpha$
and $\beta$ having values between 0 and 4 while $\gamma$ has
values between 0 and 2.

Turning to the $B^{*+}\rightarrow B^+\gamma\gamma$ amplitude, we
must consider now also the bremsstrahlung given by vertices (9.1),
(9.2). Again, we just mention here the diagrams involved, explicit
expressions being given in [22]. Firstly, there are all the
diagrams classified as in (14), except that $\mu_0$ and $\mu_+$
are interchanged. However, the major diagrams now are those
involving bremsstrahlung radiation. Let us denote the additional
diagrams by $B^{(b)} = B^{(b)}_{\rm tree} + B^{(b)}_{\rm loop}$.
$B^{(b)}_{\rm tree}$ has two classes of contributions: (1) the
chain $B^{*+}\rightarrow $``$B^{*+}$''$\gamma \rightarrow
B^+\gamma\gamma$ where the first vertex has strength $e$ and the second
$e\mu_+$; (ii) the chain $B^{*+}\rightarrow
$``$B^+$''$\gamma\rightarrow B^+\gamma\gamma$ where the first vertex
is $e\mu_+$ and the second is $e$. $B^{(b)}_{\rm loop}$ has
similarly a contribution from $B^{*+}\rightarrow \gamma
$``$B^{*+}$''$\rightarrow \gamma$``$\pi^+ B^{*0}$''$\rightarrow
\gamma\gamma B^+$ and from $B^{*+} \rightarrow $``$\pi^+ B^{*0}$''$
\rightarrow \gamma $``$B^+$''$ \rightarrow \gamma\gamma B^+$. The
radiation from the external legs is due to the (9.1), (9.2)
vertices.

In performing the calculations we used for the propagator of the
vector meson [5] $-i (g^{\mu\nu} - v^\mu v^\nu)/2[(v\cdot
k)-\Delta/4]$ and for that of the scalar meson $i/2[(v\cdot k) +
3\Delta/4]$, where $\Delta = M_{B^*} - M_{B}$. We also employed
physical masses in diagram calculations and rates, thus including
some features of the $1/M_Q$ corrections; moreover, the chiral
loops we include are themselves of order $1/M_Q$.\\

\noindent{\bf 4. ANALYSIS AND CONCLUSIONS}\\

The analysis of the various modes involves different features and
must be carried out separately. We start with the neutral decay
$D^{*0}\rightarrow D^0\gamma\gamma$. For this case, the crucial
step is to use the experimental data on the relative branching
ratios of the strong and radiative $D^{*+}, D^{*0}$ decays
$\Gamma(D^{*0}\rightarrow
D^0\pi^0):\Gamma(D^{*0}\rightarrow D^0\gamma) = (61.9\pm
2.9)\%:(38.1 \pm 2.9)\%$ and $\Gamma(D^{*+}\rightarrow
D^0\pi^+):\Gamma(D^{*+}\rightarrow D^+\gamma)=
(67.7\pm0.5)\%:(1.6\pm0.4)\%$ [1]. This allows us to relate
$\mu_0$ and $\mu_+$ to $g$ and to obtain an expression for
$\Gamma(D^{*0}\rightarrow D^0\gamma\gamma)$ in terms of $g$ at
various powers only [6]
$$\Gamma(D^{*0}\rightarrow
D^0\gamma\gamma)=[2.52\times 10^{-11}g^2 + 5.66 \times 10^{-11}
g^3$$
$$+ 4.76 \times 10^{-9}g^4 + 3.64 \times 10^{-10} g^5 + 1.53
\times 10^{-9} g^6] \ {\rm GeV} \ . \eqno(15)$$
For
$\Gamma(D^{*0}) = \Gamma(D^{*0}\rightarrow
D^0\gamma)+\Gamma(D^{*0}\rightarrow D^0\pi^0)$ we use the same
procedure for the relating of $\mu_0$ to $g$ and one obtains
$\Gamma(D^{*0}\rightarrow \ {\rm all}) = (2.02 \pm 0.12)\times
10^{-4} g^2$ GeV. This leads to our main result concerning the
measurement of the value of $g$: we have shown that the
measurement of the branching ratio $$Br(D^{*0}\rightarrow
D^0\gamma\gamma) =$$
$$ \Gamma(D^{*0}\rightarrow
D^0\gamma\gamma)/(\Gamma(D^{*0}\rightarrow D^0\gamma) +
\Gamma(D^{*0}\rightarrow D^0\pi^0)$$ given by
$$Br(D^{*0}\rightarrow D^0\gamma\gamma)=$$ $$ \frac{0.0025 +
0.057 g + 4.76 g^2 + 0.36 g^3 + 1.53 g^4}{2.02
\times 10^5} \eqno(16) $$ is a direct measurement of the
strong coupling $g$. In the expressions (15), (16) we assumed
relative positive signs for $g,\mu_0,\mu_+$ as indicated by theory
[11]. However, even if we assume negative relative signs for
various pairs of the three parameters $g, \mu_0,\mu_+$, the
changes are very small [6] since the main contribution is coming
from quadratic terms.

Since $g$ is expected to be in the range $0.25 < g < 1$, our
result (16) indicates that the branching ratio is expected to be
$0.17 \times 10^{-5}< Br(D^{*0}\rightarrow D^0\gamma\gamma)
\ < 3.3 \times 10^{-5}$. The absolute width of the two-photon decay
obtains from (15) to be $2.2\times 10^{-2} {\rm eV} <
\Gamma(D^{*0}\rightarrow D^0\gamma\gamma) < 6.7$ eV for $g$ in the
0.25-1 range.

It is of interest to remark here that for $0.1 < g < 0.7$ the main
contribution to the decay rate of $D^{*0} \rightarrow
D^0\gamma\gamma$ is given by the $g^2,g^3,g^4$ terms of (15). Out
of these terms, the more important one is the $g^4$ one, which
accounts for more than 90\% of the contribution for $g>0.25$. The
$g^4$ term gets contributions from the tree term, from the
$A^{0(1)}_{\rm loop}$ term and from the interference of the
anomaly with the $A^{02}_{\rm loop} - A^{05}_{\rm loop}$ terms.
However, the first one is by far the major contribution,
accounting for about 90\% of this term. Hence, the tree term
emerges as the major contributor to the two photon decay
$D^{*0}\rightarrow D^0\gamma\gamma$ for $g$ in the
``preferred''region $0.2 < g < 0.7$. For smaller values the
anomaly becomes competitive, while for $g$ closer to one, the
loop terms play this role. The differential decay width
$d\Gamma(D^{*0}\rightarrow D^0\gamma\gamma)/ds$ is given in [6]
for $g=0.25$ and $g=0.7$ and the role of the anomaly is visible in
the first graph.

The situation for $B^{*0} \rightarrow B^0\gamma\gamma$, which
amplitude depends also on $\mu_0$, $\mu_+$ and $g$, is
considerably more difficult to analyze since the help of measured
relative rates which we had for $D^*$ decays is unavailable here.
The branching ratio to the main decay $B^{*0}\rightarrow
B^{0}\gamma$ thus remains dependent on the three unknown couplings
in this case. Nevertheless, some help comes from the fact that the
influence of the $\mu_+$ is very small and the process may
therefore be analyzed in the reduced parameter space of $\{\mu_0,
g\}$. As shown in [6], the dependence on the assumed relative sign
of $g/\mu_0$ is however not  negligible now. Moreover, the
differential decay width does not vary enough with the variation
of $\mu_0, g$, in order to provide for a reliable determination.
It appears that the best approach is to wait for the determination
of $g$ from $D^*$ decays. Then the measurement of $B^{*0} \rightarrow
B^0\gamma$, both differential and total rate, will provide a
measurement of $\mu_0$, --- for which no other alternative exists.
Concerning the expected branching ratio, for $0.25 < g < 0.7$ and
40 eV $<\Gamma(B^{*0}) \rightarrow B^0\gamma) < 1$ KeV,
$Br\{\Gamma(B^{*0}\rightarrow
B^0\gamma\gamma)/\Gamma(B^{*0}\rightarrow B^{0}\gamma)\}$ varies
between $3.1 \times 10^{-7}$ and $1.5 \times 10^{-5}$.

Lastly, we refer to the charged decays, $D^{*+} \rightarrow
D^+\gamma\gamma$, $D^{*+}_s \rightarrow D^+_s \gamma\gamma$,
$B^{*+}\rightarrow B^+\gamma\gamma$ and let us take the latter as
an example. Now, the expected rate is much larger than in the
neutral modes, since bremsstrahlung radiation is involved. To get
an idea of the effect, just taking diagram $B_{\rm tree}^{(b)II}$
described in the previous section, one finds a branching ratio
$Br(B^{*+}\rightarrow B^+\gamma\gamma/B^{*+}\rightarrow
B^+\gamma$) of $0.7 \times 10^{-2}$ for $k_1,k_2 > 10$MeV. This
branching ratio will be mainly a function of $g$ and $\mu_+$. In
[22] a detailed analysis is presented of the effect of varying the
strength of these couplings on the branching ratio and
differential spectrum of $B^{*+}\rightarrow B^+\gamma\gamma$.
Again, the knowledge of $g$ will simplify the problem considerably.

In concluding, we remark that our calculation [6] was performed to
the leading order in chiral perturbation theory and mostly to
leading order in $1/M_Q$. Our purpose was to introduce the method,
to show how one could measure $g$ from $D^{*0}\rightarrow
D^0\gamma\gamma$ decay and to propose a possible measurement of
$\mu_{B^{*+}B^+\gamma}, \mu_{B^{*0}B^0\gamma}$ from the analysis of
$B^*\rightarrow B\gamma\gamma$ decays. Corrections to the leading
order should be performed, along the methods developed in recent
years [5,11,23], but these will not affect the qualitative
features of the method we proposed.\\

\noindent{\bf REFERENCES}
\begin{enumerate}
\item Particle Data Group, D.E. Groom et al. Eur.\ Phys.\ J.\ {\bf
C15} (2000) 1.
\item M.B. Wise, Phys.\ Rev.\ D {\bf 45} (1992) R2188; G. Burdman and J.
Donoghue, Phys.\ Lett.\ B {\bf 280} (1992) 287.
\item T.M. Yan et al., Phys.\ Rev.\ D {\bf 46} (1992) 1148; {\bf
55} (1997) 5851(E).
\item P. Cho and H. Georgi, Phys.\ Lett.\ B {\bf 296} (1992)408; B
{\bf 300} (1993) 410(E); J.F. Amundson et al., Phys.\ Lett.\ B {\bf 296}
(1992)415.
\item R. Casalbuoni et al., Phys.\ Rep.\ {\bf 281} (1997)145.
\item ACCMOR Collaboration, S. Barlag et al., Phys.\ Lett.\ B {\bf 278}
(1992)480.
\item D. Guetta and P. Singer, Phys.\ Rev.\ D {\bf 61}
(2000)054014.
\item P. Singer, Acta Phys.\ Pol. B {\bf 30} (1999) 3849.
\item H.-Y. Cheng et al., Phys.\ Rev. D {\bf 49} (1994) 5857.
\item C.G. Boyd and B. Grinstein, Nucl.\ Phys.\ B {\bf 442} (1995)
205.
\item I.W. Stewart, Nucl.\ Phys.\ B {\bf 529} (1998) 62.
\item T.N. Pham, Phys.\ Rev.\ D {\bf 25} (1982)2955; S. Nussinov
and  W. Wetzel, Phys.\ Lett.\ D {\bf 36} (1987)130; C.A. Dominguez
and N. Paver, Z.\ Phys.\ C {\bf 41} (1988)217; N. Isgur and M.B.
Wise, Phys.\ Rev.\ D {\bf 41} (1990)151.
\item E. Eichten et al., Phys.\ Rev.\ D {\bf 21} (1980) 203; S.
Godfrey and N. Isgur, Phys.\ Rev.\ D {\bf 32} (1985) 189; P.J.
O'Donnel and Q.P. Xu, Phys.\ Lett.\ B {\bf 336} (1994)113; W.
Jaus, Phys.\ Rev.\ D {\bf 53} (1996)1349;
A. Deandrea et al., Phys.\ Rev.\ D {\bf 58} (1998)034004; D.
Becirevic and A. Le Yaouannc, JHEP, 9903 (1999) 021.
\item P. Colangelo et al., Phys.\ Lett.\ B {\bf 339} (1994)151;
H.G. Dosch and S. Narison, Phys.\ Lett.\ B {\bf 368} (1996)163; P.
Colangelo and F. De Fazio, Eur.\ Phys.\ J.\ C {\bf 4} (1998)503;
F.S. Navarra et al., hep-ph/0005026.
\item V.M. Belyaev et al., Phys.\ Rev.\ D {\bf 51} (1995) 6177; A.
Khodjamirian and R. R\"{u}ckl, ``Heavy Flavours'', 2nd edition
(A.J. Buras and M. Lindner, eds.), World Scientific, Singapore
(1998); A. Khodjamirian et al., Phys.\ Lett.\ B {\bf 457}
(1999)245.
\item UKQCD Collab. (G.M. de Divitiis et al.) JHEP 9810 (1998)
010.
\item G.A. Miller and P. Singer, Phys.\ Rev.\ D {\bf 37} (1988)2564.
\item T.M. Aliev et al., Phys.\ Rev.\ D {\bf 54} (1996) 857.
\item N. Barik and P.C. Dash, Phys.\ Rev.\ D {\bf 49} (1994)299;
P. Colangelo, F. De Fazio and G. Nardulli, Phys.\ Lett.\ B {\bf 334}
(1994) 175.
\item P. Singer and G.A. Miller, Phys.\ Rev.\ D {\bf 39}
(1989)825.
\item A.K. Leibovich, A.V. Manohar and M.B. Wise, Phys.\ Lett.\ B {\bf 358}
(1995)347; {\bf 376} (1996) 332(E).
\item D. Guetta and P. Singer, to be published.
\item C.G. Boyd and B. Grinstein, Nucl.\ Phys.\ B {\bf 442}
(1995)205.
\end{enumerate}

\end{document}